# TBL-induced energy transmission into a double wall backed enclosure system computed in a cloud-based Python-FE environment

Biplab Ranjan Adhikary[a,*], Atanu Sahu[b], Partha Bhattacharya[a]

[a]Jadvapur University, Kolkata, India 700 032

[b]National Institute of Technology Silchar, Silchar, India 788 010

[*]bradhikary.civil.rs@jadavpuruniversity.in

A B S T R A C T

We propose a fully coupled numerical model to predict turbulent boundary layer (TBL) induced energy transmission behavior for a double-wall backed enclosure system in a finite element (FE) framework computed in cloud-based Python environment. Goody's single point wall-pressure spectrum and Corcos spatial correlation function are used to generate the TBL cross-power spectra. Mindlin's first order shear deformation model is considered for the panels and a fully coupled TBL-structure-acoustic model is developed using the FE approach to predict the acoustic power level inside the enclosure for variable gap distance between the panels. The model is developed in a way to capture the contribution of orthotropic lamina sequence, frequency-dependent structural damping, and stiffening orientation in predicting the energy transmission into a double-wall backed enclosure. Thus, a new numerical model is presented that enables the designers with more precise energy transmission quantification with greater flexibility in terms of the number of panel leaves, geometry, and boundary conditions of the enclosure system, backed by double wall made of isotropic or orthotropic laminates.

*Keywords: double-wall backed enclosure, finite element, turbulent boundary layer, energy transmission, Python cloud*

## 1. Introduction

Turbulent boundary layer (TBL) induced energy transmission estimation into an acoustic enclosure backed by double wall is performed by the researchers in the wave number-frequency domain using the analytical modal expansion technique [1]. However, the analytical method essentially restricts one to incorporate complex panel geometry and boundary conditions. Therefore, in the present work a FE-based numerical framework is developed and presented which predicts TBL-excited energy transmission into a double-wall backed acoustic enclosure.

The computation of the entire coupled TBL-structure-acoustics system is performed in cloud-based Python environment and astonishingly greater time efficiency is achieved in comparison to MATLAB.

A zero-pressure gradient TBL is considered in this model and the single-point wall-pressure spectrum is calculated using the Goody model [2]. Modified Corcos spatial coherence function [3-4] is used to generate pressure cross-PSD. Coupled system transfer function matrices are estimated using component transfer functions and developed extended coupling equations. Finally, TBL-induced energy transmission into the enclosed cavity is estimated in form of the point pressure PSD.

## 2. Mathematical formulation

### 2.1. Aerodynamic modelling

The turbulent flow field is considered to be homogeneous and stationary. Single-point TBL wall-pressure spectrum ($\Phi_p$) is calculated using Goody model [2] as reported by Caiazzo et al. [1]. The Goody model used to calculate wall-pressure spectrum, is given in Eq. (1).

$$\Phi_p = \frac{3(2\pi f \tau_\omega)^2 (\frac{\delta}{U})^3}{\left(\left(\frac{2\pi f \delta}{U}\right)^{\frac{3}{4}} + 0.5\right)^{3.7} + \left(1.1 R_T^{-0.57} \frac{2\pi f \delta}{U}\right)^7} \tag{1}$$

where $\delta$ is the TBL thickness, $\tau_\omega (= \rho U_\tau^2)$ is the wall shear stress with $\rho$ as density of air and $U_\tau$ friction velocity. $R_T = \frac{U_\tau^2 \delta}{U_\infty \nu}$ is the Reynolds number dependent factor. $U_\infty$ is the free-stream velocity and $\nu$ is the kinematic viscosity of air.

The excited panel (outer panel) is first discretized using a 2D grid and the wall-pressure cross power spectral density (cross-PSD) over all the grid points are calculated using single-point wall spectra and the coherence function as per the modified Corcos model [4],

$$\Phi_{pp}(x_\mu, x_\nu, \omega) = \sqrt{\Phi_p(x_\mu, \omega)\Phi_p(x_\nu, \omega)} \Gamma(\xi_1, \xi_3, \omega) \tag{2}$$

where, the spatial correlation function is expressed] as,

$$\Gamma(\xi_x, \xi_z, \omega) = \left(1 + \alpha_x \left|\frac{\omega \xi_x}{U_c}\right|\right) e^{-\alpha_x \left|\frac{\omega \xi_x}{U_c}\right|} e^{i\frac{\omega \xi_x}{U_c}} e^{-\alpha_z \left|\frac{\omega \xi_z}{U_c}\right|} \tag{3}$$

$\xi_x$ and $\xi_z$ are separation vectors between two points in the streamwise and cross-stream directions, respectively. Corcos model constants are taken as $\alpha_x = 0.11$, $\alpha_z = 0.70$. [4]

## 2.2. Turbulence-structure coupling

The structural panel(s) is/are discretized using 4-node iso-parametric elements and the discretization is so done that the cavity (enclosure and/or gap) pressure FE grid(s) and the structural FE grids coincide and properly mapped.

### 2.2.1. Panel-enclosure system

The enclosed cavity in the single-leaf panel system is modelled using 8-node octahedral FE with pressure DOF. The external force-induced response of the entire system can be expressed as,

$$Y(\omega) = H(\omega)X(\omega) \tag{4}$$

Here, $H(\omega)$ and $X(\omega)$ are the frequency-dependent transfer function and forcing function, respectively.

$$H(\omega) = \begin{bmatrix} H_{11} & H_{12} \\ H_{21} & H_{22} \end{bmatrix}^{-1} \tag{5}$$

Designation: 1-panel; 2-enclosure

The uncoupled and coupled transfer functions $(H_{ij})$ are derived from the dynamic equations of the panel and cavity in modal domain, as detailed in Eq. (5) and Eq. (12) of Ghosh and Bhattacharya [5].

Subscripts, $i = j$ represent auto coupling, and $i \neq j$ represent cross-coupling. The TBL forces act on the panel only, and hence the force vector can be written as,

$$X(\omega) = \begin{Bmatrix} F_{tbl} \\ 0 \end{Bmatrix} \tag{6}$$

On assembling Eq. (5) and Eq. (6) as described in Eq. (4) and rearranging, one obtains the response function of the panel ($H_w$) and the enclosed cavity ($H_e$) for the coupled system,

$$H_w = (H_{11} - H_{12}H_{22}^{-1}H_{21})^{-1} \tag{7}$$

$$H_e = -H_{22}^{-1}H_{21}H_w \tag{8}$$

The modal response functions are so arranged that they can be solved for any number of modes for panel 'a', panel 'b' and cavity. Once, the response functions are obtained in a coupled system, the panel displacement PSD ($S_{ww}$) and enclosure pressure PSD ($S_{ee}$) can be calculated as, [6]

$$S_{ww}(\omega) = H_w^*(\omega)S_{tbl}(\omega)H_w^T(\omega) \tag{9}$$

$$S_{ee}(\omega) = H_e^*(\omega) S_{tbl}(\omega) H_e^T(\omega) \tag{10}$$

Here, $S_{ww}(\omega)$ is the panel displacement PSD, $S_{tbl}(\omega)$ is the TBL cross PSD of the force, calculated as,

$$S_{tbl}(\omega) = A_\mu \Phi_{pp} A_\nu \tag{11}$$

$A_\mu$ and $A_\nu$ are the elemental areas around the nodes $\mu$ and $\nu$.

*2.2.2. Panel-gap-panel-enclosure system*

The mathematical framework is an extension of the 'panel-enclosure' system, but with more complex fully coupled behavior accounting the TBL excitation effect on an acoustic enclosure backed by double-walled panel system that are separated by a gap. The gap and the enclosure both are modelled using 8-node octahedral FE with pressure DOF. The external force-induced response of the entire structural system can be expressed as Eq. (4)

$$\text{Where } H(\omega) = \begin{bmatrix} H_{11} & H_{12} & H_{13} & H_{14} \\ H_{21} & H_{22} & H_{23} & H_{24} \\ H_{31} & H_{32} & H_{33} & H_{34} \\ H_{41} & H_{42} & H_{43} & H_{44} \end{bmatrix}^{-1} \tag{12}$$

Designation: 1-skin panel (panel 'a'); 2-trim panel (panel 'b'); 3-gap; 4-enclosure

The transfer functions $(H_{ij})$ are derived from the dynamic equations of the two panels, gap and enclosure in modal domain, as detailed in Eq. (5), Eq. (6) and Eq. (12) of Ghosh and Bhattacharya [5].

Subscripts, $i = j$ represent auto coupling, and $i \neq j$ represent cross-coupling. As there is no direct connection between a) the two panels, b) skin panel-enclosure, and c) gap-enclosure, $H_{12} = H_{21} = H_{14} = H_{41} = H_{34} = H_{43} = 0$. The TBL forces act on the skin panel only, and hence the force vector can be written as,

$$X(\omega) = \begin{Bmatrix} F_{tbl} \\ 0 \\ 0 \\ 0 \end{Bmatrix} \tag{13}$$

On putting Eq. (12) and Eq. (13) in Eq. (4) and rearranging, one can obtain the response function of the panels ($H_{w,a}$ and $H_{w,b}$) and the cavity ($H_p$) for the fully coupled system. But, there are two challenges:-

a) Decoupling of the equations should be performed in such a way that initially only the coefficients of the $H_{ii}$ form are inverted, as the other coefficients are not necessarily square matrices to keep the generic fabric of the entire matrix system intact. As the solution proceeds, one can find combinations of matrix coefficients arranged together in square forms which can then be inverted.

b) Multiplication of matrix coefficients are to be done keeping their dimensions in mind. As all the four systems (two panels, two cavities) may have different number of modes accounted, violation of this rule otherwise will lead to further errors.

Finally, the solution developed in the present model takes the following form:

$$H_{w,b} = -[H_{22}^{-1}H_{23}H_{33}^{-1}H_{32} - (I - H_{22}^{-1}H_{24}H_{44}^{-1}H_{42})(I - H_{33}^{-1}H_{31}H_{11}^{-1}H_{13})]^{-1}H_{22}^{-1}H_{23}H_{33}^{-1}H_{31}H_{11}^{-1} \quad (14)$$

$$H_g = (I - H_{33}^{-1}H_{31}H_{11}^{-1}H_{13})^{-1}H_{33}^{-1}(H_{31}H_{11}^{-1} + H_{32}H_{w,b}) \quad (15)$$

$$H_e = -H_{44}^{-1}H_{42}H_{w,b} \quad (16)$$

$$H_{w,a} = H_{11}^{-1}(I - H_{13}H_g) \quad (17)$$

The present formulation works perfectly all right when the number modes of panel 'b' and gap are equal. If these two are required to be different, the solution can be obtained by taking other possible ways also. In the coupled system, the panel displacement PSDs ($S_{ww,a}$ & $S_{ww,b}$) and gap and enclosure pressure PSDs ($S_{gg}$ & $S_{ee}$) can be calculated as,

$$S_{ww,a}(\omega) = H_{w,a}^*(\omega)S_{tbl}(\omega)H_{w,a}^T(\omega) \quad (18)$$

$$S_{ww,b}(\omega) = H_{w,b}^*(\omega)S_{tbl}(\omega)H_{w,b}^T(\omega) \quad (19)$$

$$S_{gg}(\omega) = H_g^*(\omega)S_{tbl}(\omega)H_g^T(\omega) \quad (20)$$

$$S_{ee}(\omega) = H_e^*(\omega)S_{tbl}(\omega)H_e^T(\omega) \quad (21)$$

The panel and cavity responses can be used to estimate cavity pressure, skin panel response, etc. But as the present work is focused on the estimation of the transmitted sound power into the enclosure, the enclosure pressure PSD matrix, $S_{ee}(\omega)$ is considered, that consists of pressure PSD at all the enclosure nodes. Subsequently, point pressure PSD and average enclosure pressure PSD can be estimated.

## 4. Results and discussion

*4.1. Energy transmission into double-wall backed enclosure: validation*

The present developed model is validated by the analytical work reported by Caiazzo et al. [1]. The model properties of the panel-gap-panel-enclosure system is considered same as Table 1 in p. 168 of Caiazzo et al. [1]. Panel 'a' is 2mm and panel 'b' is considered 3mm thick. FE meshing for flexible panels is adopted as to keep the element size well below the wavelength ($\Delta x < \lambda/3$) of the plate bending wave, in order to account for the convective part of the pressure fluctuations [7]. The in-vacuo free vibration analysis for structural panels, and rigid boundary free vibration analysis for the gap and enclosure are performed using in-house Python codes. These yields the frequency and mode number data, for all the systems, which are presented in Table 1.

**Table 1** First *eight* frequencies (Hz) upto 1000Hz

| Structural Panel | | Cavity | | |
| --- | --- | --- | --- | --- |
| Panel 'a' | Panel 'b' | Gap | Enclosure | Coupled |
| 108 | 162 | 429 | 429 | 134 |
| 225 | 337 | 572 | 431 | 202 |
| 317 | 474 | 715 | 572 | 322 |
| 421 | 630 | 858 | 608 | 419 |
| 433 | 649 | | 715 | 444 |
| 628 | 941 | | 716 | 592 |
| 667 | 999 | | 834 | 632 |
| 696 | | | 858 | 698 |

In order to perform robust calculations solving all the matrix equations in FE framework, cloud-based Python environment is deployed. 24 modes for panel 'a', 20 modes for panel 'b' and gap, and 48 modes for enclosure are considered for validation and all the subsequent calculations. The TBL-excited enclosure pressure PSD at a given point, referring Fig. 19 of Caiazzo et al., 2018 [1], is estimated using the developed FE model, and presented in Fig. 1. The estimated result is found to be in excellent agreement with the analytical result reported by Caiazzo et al. [1]

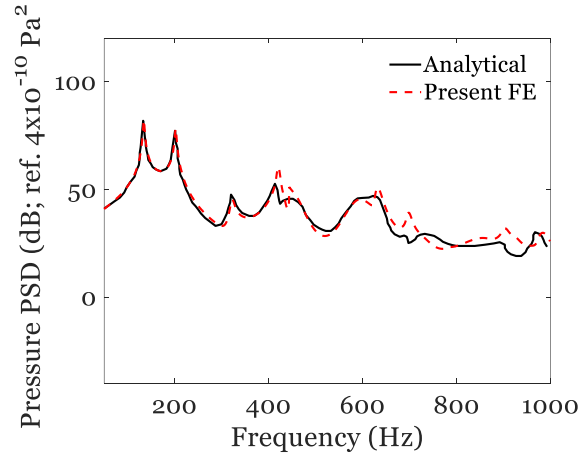

**Fig. 1.** Enclosure pressure PSD at $(L_x/3, L_y/3, -2L_z/3)$

*4.2. Enclosure pressure with variable gap distance: case study*

Once the FE model is validated, next enclosure pressure PSD at the same location is estimated keeping all the conditions as previous. The only variable is the the gap distance between two panels. Results are obtained for three different gap distances, 0.019m, 0.038m and 0.076m, and presented in Fig. 2.

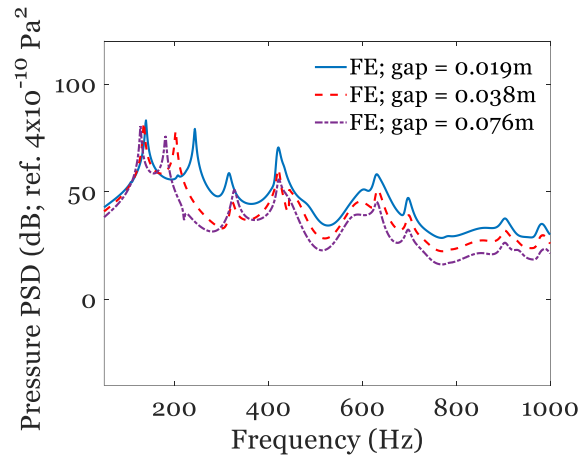

**Fig. 2.** Pressure PSD at $(L_x/3, L_y/3, -2L_z/3)$ for different gap distance

It is observed that, (1) lowest gap distance (0.019) produces maximum point pressure into the enclosure, and in case of highest gap distance (0.076m) the transmitted pressure into enclosure is minimum, (2) increasing the gap distance from 0.038m to 0,076m is generally reducing the enclosure pressure, but in the 0-400Hz region, this is not always true, especially in the 300-450Hz range. 3rd (322Hz) and 4th (422Hz) coupled modes may be responsible for this changed behavior, (3) there is frequency shift in the coupled system, but the 2nd fundamental frequency shift for lowest gap (0.019m) is quite significant. This is due to reduction in gap the 2nd fundamental mode of panel 'a' is dominating the system.

## 5. Conclusion

A fully coupled numerical FE model is developed to estimate TBL-induced energy transmission into an enclosure backed by double-wall, and solved using cloud-based Python environment. The enclosure pressure prediction of the present FE model is found to be excellent. Subsequently, a case study is performed with variable gap distance. It is observed that 0.019m gap distance is transmitting highest level of energy of all the cases. Increasing the gap distance from 0.038m to 0.076m is not exhibiting transmission reduction in the 0-400Hz, as was expected. The present model can incorporate complex structural geometry and boundary conditions for the panels made of isotropic or orthotropic laminates.

**CRediT authorship contribution statement**

**Biplab Ranjan Adhikary:** Conceptualization, simulation, validation, manuscript preparation, review & editing. **Atanu Sahu:** Conceptualization, review and editing. **Partha Bhattacharya:** Conceptualization, review and editing.

**Declaration of Competing Interest**

The authors declare that they have no known competing financial interests or personal relationships that could influence the work reported in this paper.